# The neural correlates of logical-mathematical symbol systems processing resemble that of spatial cognition more than natural language processing

Yuannan Li[1], Shan Xu[2]*, Jia Liu[1]

*: corresponding author

*1 Department of Psychology & Tsinghua Laboratory of Brain and Intelligence, Tsinghua University, Beijing, 100086, China*

2 *Faculty of Psychology, Beijing Normal University, Beijing, 100875, China*

Address correspondence to Shan Xu, Faculty of Psychology, Beijing Normal University, Beijing, 100875, China. Email: shan.xu@bnu.edu.cn

**Abstract:** The ability to manipulate logical-mathematical symbols (LMS), encompassing tasks such as calculation, reasoning, and programming, is a cognitive skill arguably unique to humans. Considering the relatively recent emergence of this ability in human evolutionary history, it has been suggested that LMS processing may build upon more fundamental cognitive systems, possibly through neuronal recycling. Previous studies have pinpointed two primary candidates, natural language processing and spatial cognition. Existing comparisons between these domains largely relied on task-level comparison, which may be confounded by task idiosyncrasy. The present study instead compared the neural correlates at the domain level with both automated meta-analysis and synthesized maps based on three representative LMS tasks, reasoning, calculation, and mental programming. Our results revealed a more substantial cortical overlap between LMS processing and spatial cognition, in contrast to language processing. Furthermore, in regions activated by both spatial and language processing, the multivariate activation pattern for LMS processing exhibited greater multivariate similarity to spatial cognition than to language processing. A hierarchical clustering analysis further indicated that typical LMS tasks were indistinguishable from spatial cognition tasks at the neural level, suggesting an inherent connection between these two cognitive processes. Taken together, our findings support the hypothesis that spatial cognition is likely the basis of LMS processing, which may shed light on the limitations of large language models in logical reasoning, particularly those trained exclusively on textual data without explicit emphasis on spatial content.

**Keywords:** logical-mathematical symbols processing, natural language processing, spatial cognition, language of thought, large language model

# 1. Introduction

Throughout its relatively brief history, humans have devised and mastered numerous logical-mathematical symbol (LMS) systems such as logical reasoning and calculation, reflecting our cultural and technological evolution. This trend continues with the creation of novel LMS systems with emerging knowledge domains such as programming, showcasing the remarkable adaptability of the human mind to increasingly complex cognitive challenges. In addition, the use of LMS systems to tackle abstract problems is attracting further interest, particularly for its potential to bridge the intellectual gap between artificial intelligence (AI) and human cognition. Despite its diverse applications and functionalities, LMS processing is theorized to originate from a single foundational system, namely the logical-mathematical intelligence, as proposed by Gardner's multiple intelligences theory (1983). However, the exact cognitive functionality of this foundational system remains largely elusive. Especially, considering that known LMS-related behaviors have emerged relatively recently, within the last 6000 years (Pande, 2010), it is unlikely that a completely new set of neural facilities evolved specifically for this purpose. Accordingly, LMS processing may have evolved by repurposing phylogenetically more ancient cognitive systems through neuronal recycling (Dehaene, 2005; Dehaene & Cohen, 2007; Fedorenko et al., 2019; Kazanina & Poeppel, 2023).

The primary systems thought to underlie LMS processing are natural language processing and spatial cognition (Dehaene & Cohen, 2007; Fedorenko et al., 2019). In terms of form, LMS systems bear an ostensible resemblance with natural language (Dehaene et al., 2022; Fedorenko et al., 2019). This similarity is especially evident in the syntax-like operations of LMS, which are thought to parallel those in language processing (Fedorenko et al., 2019). For example, the verbal reasoning approach, which views deductive reasoning, a representative LMS task, as a transformation of verbal information, relies on processes similar to those in language comprehension and generation (Krumnack et al., 2011; Polk & Newell, 1995). However, empirical evidence regarding the association between language and LMS processing is inconclusive, both in terms of correlation in behavioral performance (Fedorenko & Varley, 2016; LeFevre et al., 2010; Prat et al., 2020; Shearer & Karanian, 2017) and identification of common neural correlates (Amalric & Dehaene, 2016; Castelhano et al., 2021; Coetzee et al., 2022; Fedorenko & Varley, 2016; Floyd et al., 2017; Houdé & Tzourio-Mazoyer, 2003; Ivanova et al., 2020; Y.-F. Liu et al., 2020; Maruyama et al., 2012; Monti et al., 2009, 2012; Prat et al., 2020; Wang et al., 2020), implying the complexity of the relationship between LMS and language systems. In addition, the comparison is limited to a handful of representative tasks, such as prose cloze versus code function judgment, sentence reading versus mental programming, and syllable concatenation versus number addition (Endres, Karas, et al., 2021; Xu et al., 2021; Zago et al., 2008), and the specifics of corresponding task settings risk confounding the test of commonality (but see Ikutani, Itoh, et al., 2021 for a comparison between programming comprehension with the meta-analysis map of language, and Castelhano et al., 2021 for a comparison between program understanding with the meta-analysis results of reading).

An alternative hypothesis proposes that LMS processing is built upon spatial cognition, which views the processing of the properties and relations among non-spatial LMS elements dependent on spatial abilities (Dehaene & Cohen, 2007; Lin & Dillon, 2023). For instance, the mental model theory argues that reasoning is achieved by manipulating reasoning components in a visuospatial workspace, rather than a linguistic medium (Johnson-Laird, 2001). Similarly, the mental simulation theory posits that the ability to navigate physical environments underpins the simulation processes used to arrive at solutions in a logical-mathematical space (Barsalou, 2008; Hawes & Ansari, 2020). Supporting these theoretical models is behaviour and neural evidence linking LMS processing abilities, such as numerical, arithmetical, and geometric abilities (Houdé & Tzourio-Mazoyer, 2003; Wei et al., 2012; Xie et al., 2020), logical reasoning (Goel, 2007; Houdé & Tzourio-Mazoyer, 2003; Knauff et al., 2003), programming (Helmlinger et al., 2020; Huang et al., 2019; Sharafi et al., 2021), computational thinking (Román-González et al., 2017; Xu et al., 2021), and other STEM domains (Wai et al., 2009; Khine, 2017; Shea et al., 2001) to spatial ability. This suggests that spatial cognition may play a critical role in facilitating LMS processing, offering an important perspective alongside the language-based approach. However, similar to the situation with the language hypothesis, only a handful of specific LMS and spatial tasks were utilized in the comparison (Endres, Fansher, et al., 2021; Endres, Karas, et al., 2021; Huang et al., 2019; Zago et al., 2008). For example, in our previous study (Xu et al., 2021), to infer the cognitive substrates of computational thinking, we broke up the neural correlates of mental programming according to their varied pattern of neural resemblance with three other LMS tasks and a representative visuospatial task (mental folding), and we observed that the part showing resemblance between mental programming and other LMS tasks does not necessarily show high resemblance with mental folding. This leaves open a question of whether the hypothetical neural resemblance between LMS and spatial processing exists above task variation, being evident at the domain level, and whether such neural resemblance is constantly larger than the LMS-language resemblance across tasks.

To examine these two hypotheses at the domain level, here we compared the involvement of cortical regions associated with language processing and spatial cognition in LMS processing, respectively. Our rationale, grounded in the concept of neuronal recycling (Dehaene, 2005; Dehaene & Cohen, 2007), is that the basis of LMS processing is more likely to be the cognitive function whose cortical regions are predominantly engaged in LMS processing. The present study presented a novel attempt at direct and quantitative domain-level examination of the two hypotheses.

Instead of comparing the activation of specific tasks, we ran large-scale automated meta-analyses to identify the cross-task neural correlates of language processing and spatial cognition, respectively. For LMS processing, with few dedicated studies, we operationalized it by its three core cognitive units as defined in logical-mathematical intelligence (Gardner, 1983), namely, calculation, logical reasoning, and problem-solving, and compiled a synthesized activation map through a conjunctive analysis of whole-brain univariate activations based on inhouse data of three representative tasks of each core unit: reasoning, calculation, and mental programming from our previous work for another purpose (Xu et al., 2021).

With these meta-analysis results, we quantified and compared the domain-level overlap of LMS processing with language processing and spatial cognition, respectively. We indeed observed greater domain-level resemblance of LMS with spatial processing than with language processing, the cross-task homogeneity of which was further validated by multivariate activation pattern analysis**.** Finally, we used clustering analysis to explore the hierarchical structure among the neural correlates of LMS processing and its functional bases. These analyses collectively revealed a notable link between LMS processing and spatial cognition, supporting the hypothesis that the root of LMS processing lies in spatial cognition.

## 2. Results

### 2.1. Univariate analysis revealed more overlap of LMS processing with spatial cognition

We first investigated whether LMS processing shared neural activation to a greater extent with language processing or spatial cognition. To identify the neural correlates intrinsic to LMS processing, distinct from those linked to task-specific properties, we conducted a conjunctive analysis using fMRI data from three representative tasks: reasoning, calculation, and mental programming (see Methods 4.2.2) (Figure 1a, middle). This synthesis allowed the generation of an activation map associated with LMS processing embedded across these three tasks. The resulting map revealed bilateral activation, in the frontal lobe (the middle frontal gyrus, the inferior frontal gyrus), the supplementary motion area spanning to the cingulum, the insula, the posterior parietal cortex, the precuneus, and the inferior temporal gyrus (Supplemental Figure 1, Supplemental Table 1).

To map the neural substrates of language processing and spatial cognition, we leveraged large-scale automated meta-analyses to generate cross-task activation maps for each (see Methods 4.1) (Figure 1a, left and right). Language processing activates a left-lateralized network, including the precentral and the inferior frontal gyri, the supplementary motion area, and the superior and middle temporal gyri (Supplemental Figure 2, Supplemental Table 2). In contrast, spatial cognition recruits the bilateral superior frontal gyrus, the opercular part of the inferior frontal gyrus, the supplementary motion area, the insula, the superior parietal lobule spanning to the middle occipital gyrus, the precuneus, and the inferior occipital gyrus spanning to the fusiform (Supplemental Figure 3, Supplemental Table 3). These activation patterns are consistent with findings from previous reviews (Cona & Scarpazza, 2019; Price, 2012).

Figure 1c shows the activation overlap among LMS processing, language processing, and spatial cognition. For a full list of the overlapping clusters, refer to Supplementary Table 4. Visual inspection shows that the activation map of LMS processing overlapped with that of language processing mainly in the left inferior frontal gyrus and the left supplementary motion area. In contrast, overlap with spatial cognition is mainly in the bilateral posterior parietal cortex, the inferior and the middle frontal gyri, the right supplementary motion area, the right insula, and the right precuneus. Critically, there is substantial co-activation between LMS processing and spatial cognition, and to a somewhat lesser extent, between LMS processing and language processing. In the neural activation of LMS processing, the voxels overlapping with spatial cognition (n = 1694) amount to 2.7 times the number of voxels overlapping with language processing (n = 634). This difference is also reflected in the dice coefficient, where the coefficient between spatial cognition and LMS processing (0.255) was larger than that between language processing and LMS processing (0.113), suggesting a greater similarity in the univariate activation maps of spatial cognition and LMS processing.

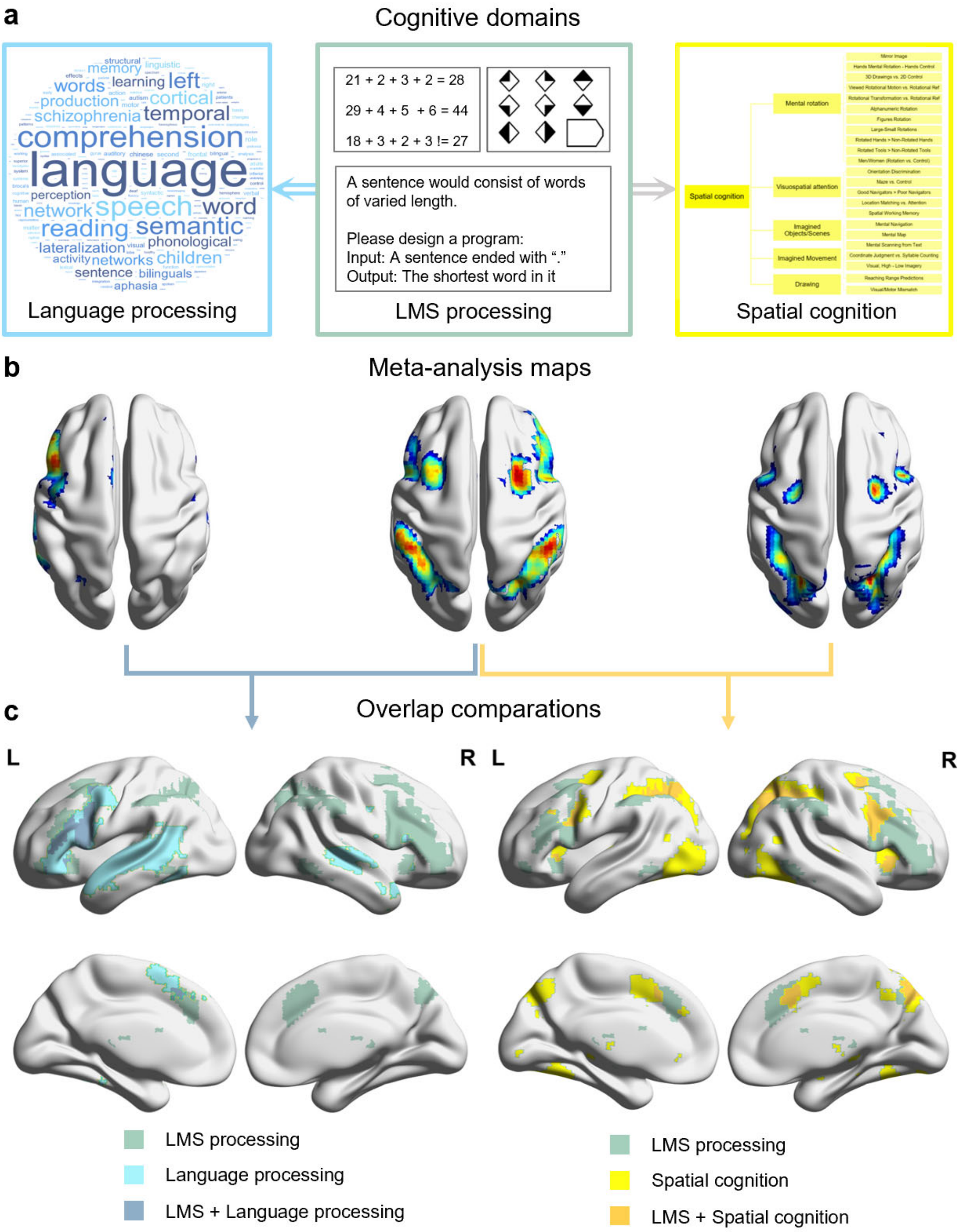

Figure 1. The comparison between neural correlates of LMS processing, language processing, and spatial cognition. (a) The generation of the domain-level neural correlate maps. LMS processing (middle): Image-based meta-analysis by representative tasks, i.e., reasoning, calculation, and mental programming. Language processing (left): automated meta-analysis by Neurosynth based on large-scale text

statistics. The word cloud showed the composition of the words from the titles of the included papers, excluding neuroimaging terminology such as “fMRI”. Spatial cognition (right): automated meta-analysis by Brainmap based on the established categorization (Fox et al., 2005). The tree graph displayed the experimental paradigms and contrasts included. (b) The dorsal view of the resultant domain-level maps. The warmer color indicated a stronger likelihood of activation in the meta-analysis. All the maps were FDR corrected at $p = 0.01$. (c) The overlap between the domain-level maps of LMS processing and either language processing or spatial cognition.

To replicate these domain-level findings and statistically compare the resemblance of LMS processing to either language processing or spatial cognition, we calculated the subject-wise overlap between each subject’s LMS synthesized map and their fMRI activation in representative language and spatial tasks, sentence processing, and mental folding, respectively (see methods 4.2.2) (Figure 2a, left and right). All the subject-wise maps were thresholded by a voxel-level threshold at $p = 0.001$. Consistent with our observations from the domain-level analysis, the dice coefficients between spatial cognition and LMS processing were significantly higher than those for language processing and LMS processing ($t(19) = 8.825$, $p < 0.001$, *Cohen’s d* = 2.025) (Figure 2b, left). This result illustrates the greater similarity in neural activation of LMS processing and spatial cognition compared to that between LMS processing and language processing.

Given that our synthesized LMS activation map was derived from three distinct representative LMS tasks, we further asked whether our findings were consistent across each task. To this end, we compared the group-level activation map of each LMS task (corrected by FDR at $p = 0.01$) with the domain-level language and spatial processing maps, and further statistically tested the difference between the task-wise LMS activations and those of sentence processing and mental folding.

In the reasoning task, overlap with spatial cognition was mainly in the bilateral posterior parietal cortex, the inferior frontal gyrus, the right middle frontal gyrus, the right insula, and the left precuneus, while overlap with language processing was mainly in the left inferior and middle frontal gyri, and the left supplementary motion area (Supplemental Figure 4). The overlap of reasoning with spatial cognition doubled the size of its overlap with language processing. Consistently, the dice coefficient for reasoning with spatial cognition (0.280) was larger than that with language processing (0.170). This pattern was confirmed by a significant difference between corresponding subjective dice coefficients ($t(19) = 5.770$, $p < 0.001$, *Cohen's d* = 1.324) (Figure 2b, right).

Similar patterns were observed for calculation and mental programming tasks. In the calculation task, overlap with spatial cognition was mainly in the bilateral inferior parietal lobule, the supplementary motion area, the left middle frontal gyrus, the opercular part of the right inferior frontal gyrus, and the right insula, while that with language processing was relatively limited, predominantly in the opercular part of the left inferior frontal gyrus (Supplemental Figure 5). Overlap between calculation and spatial cognition was 1.8 times larger than its overlap with language processing, consistent with the comparison of dice coefficients (with spatial cognition: 0.149 versus with language processing: 0.053), indicating a more substantial resemblance between the neural correlates of calculation and spatial cognition. This pattern was confirmed by a significant difference between corresponding subjective dice coefficients ($t(19) = 3.477$, $p < 0.01$, *Cohen's d* = 0.798) (Figure 2b, right).

In the mental programming task, overlap with spatial cognition was mainly in the left inferior parietal lobule and the left precuneus, while that with language processing was mainly in the left precentral gyrus and the triangular part of the left inferior frontal gyrus (Supplemental Figure 6). The overlap between mental programming and spatial cognition is more than 10 times the size of the overlap between mental programming and language processing, consistent with the larger programming-spatial than programming-language dice coefficients at both the group level (with spatial cognition: 0.065 versus language processing: 0.044) and subject level ($t(19) = 5.215$, $p < 0.001$, *Cohen's d* = 1.196) (Figure 2b, right), suggesting that the neural activation in mental programming was more similar to spatial cognition than language processing.

In summary, we found substantially greater overlap in activation maps of LMS processing with spatial processing, especially in the posterior parietal cortex and inferior frontal gyrus, supporting the spatial origin hypothesis.

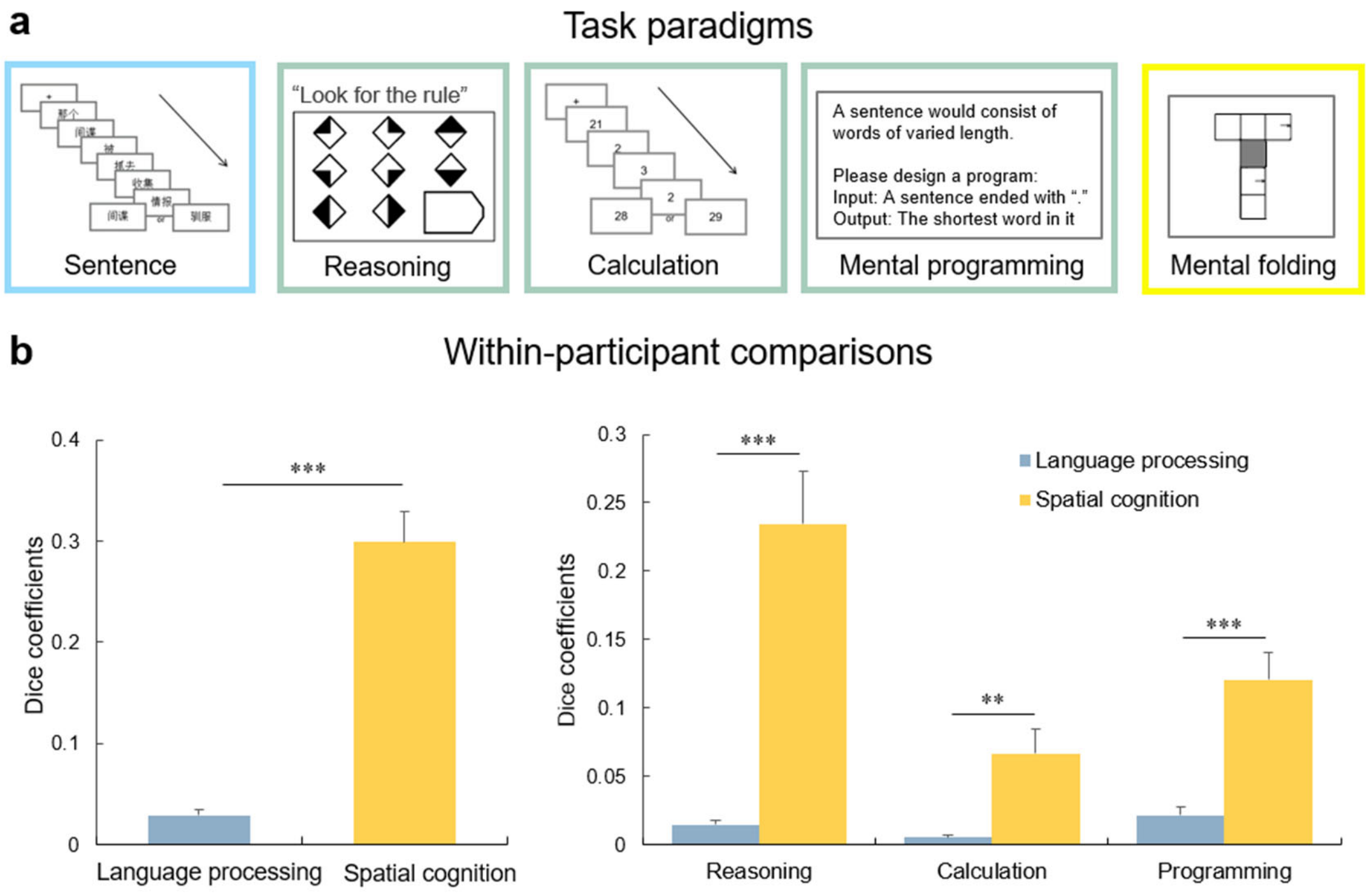

Figure 2. Task-level comparison between the univariate activation of LMS processing, language processing, and spatial cognition. (a) The representative tasks for each domain. The LMS processing tasks (middle): reasoning, calculation, and mental programming; the language processing task (left): sentence processing; the spatial cognition task (right): mental folding. (b) The subject-wise LMS-spatial and LMS-language dice coefficients (left) and the corresponding dice coefficients after breaking LMS into separate representative tasks (right). All the subject-level activation maps were voxel-level thresholded at $p = 0.001$. The error bar denotes the standard error (SE). The asterisks denote the significance of the difference between the dice coefficients. *$p$<.05, **$p$<.01, ***$p$<.001.

## 2.2. Similarity in activation patterns between LMS processing and spatial cognition

While the univariate analysis above revealed that LMS processing shared more similarities with spatial cognition in terms of activation maps overlap, it is important to note that a single cortical region can support distinct cognitive processes (Duncan, 2010; Woolgar et al., 2011, 2016). Particularly, the frontal and parietal cortex were known to be functionally heterogeneous and showed substantial cross-individual variation (Blank & Fedorenko, 2017; Braga et al., 2019; Frost & Goebel, 2012; Ivanova et al., 2020; Tahmasebi et al., 2012; Vázquez-Rodríguez et al., 2019). Therefore, anatomical overlap in activation between LMS processing and spatial cognition in these regions does not necessarily imply functional similarity. To further test the hypothetical functional association between LMS and spatial processing, we examined whether the multivariate activation pattern of LMS processing more closely resembled that of spatial cognition than language processing. For an unbiased comparison, we deliberately chose cortical regions activated by both language processing and spatial cognition. This region of interest (ROI), delineated as the overlap between the corresponding meta-analysis activation maps, was left-lateralized, encompassing the precentral gyrus, the inferior frontal gyrus, the supplementary motor area, and the posterior parietal cortex, as well as the bilateral superior temporal gyrus (Supplemental Figure 7, Supplemental Table 5).

The similarity between domains was quantified by examining the likelihood of the multivariate activation patterns of LMS processing being classified as those of the representative spatial task rather than those of the representative language task in the ROI. To do this, we trained a support vector machine (SVM) for each subject, using it to classify the multivariate patterns of language processing, indexed by the sentence processing task, from those of spatial cognition, indexed by the mental folding task (Figure 3a) with leave-one-out cross-validation. Having established the boundary surface that divides multivariate patterns into two distinct classes, language processing and spatial cognition (Figure 3b), we then applied this classifier to the untrained multivariate patterns of LMS processing, indexed by the three representative tasks (Figure 3d). The underlying rationale is that if the ROI represented LMS in a manner more similar to spatial codes than to linguistic elements, then the classifier would more frequently classify LMS processing as spatial cognition than language processing (Figure 3e).

As expected, the spatial-linguistic SVM classifier demonstrated an accuracy close to 100% in distinguishing the multivariate patterns of language processing from those of spatial cognition across all subjects (Figure 3c), demonstrating the effectiveness of the classifier. Critically, when applying this classifier to the multivariate patterns of LMS processing in the ROI, the likelihood of LMS processing being classified as spatial cognition was 65%, significantly surpassing the 35% likelihood of being classified as language processing ($t(19) = 3.127$, $p = 0.006$, *Cohen's d* = 0.717) (Figure 3f). That is, in regions recruited by both language processing and spatial cognition, the activation patterns elicited by LMS processing were more similar to those of spatial cognition than to language processing.

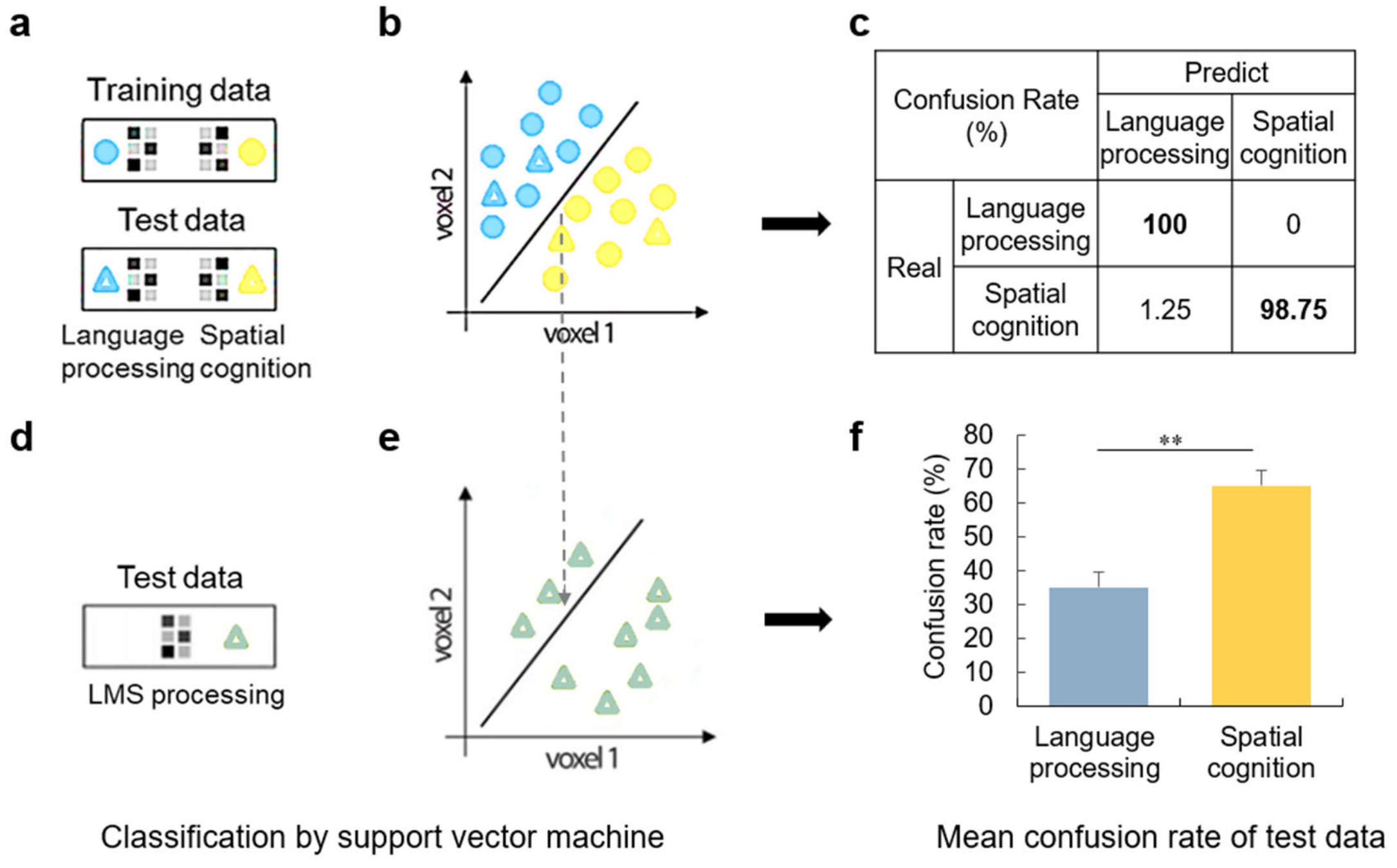

Figure 3. ROI analysis of the multivariate pattern similarity between LMS processing and spatial cognition (or language processing). (a-c) The schematic illustration of the SVM trained to classify language processing and spatial cognition. (a) illustrates the composition of the training (dots) and the testing (triangles) data. Blue denotes language processing and yellow spatial cognition. The gray matrices in the box are an illustration of the training and the testing data, i.e., the individual activation patterns during corresponding task runs in the ROI. (b) illustrates the boundary surface for the two target classes (the line). The dots above the line would be classified as language processing, and the dots below the line as spatial cognition. (c) shows the resultant confusion matrix averaged across participants, whose rows correspond to the real class of the data, and the columns represent the predicted class by the SVM. (d-f) The schematic illustration of the classification of LMS processing activation patterns. The data of LMS processing (green triangles, d) was classified by the identical SVM (e)

from b of the same subject. The dots above the line would be classified as language processing, and the dots below the line as spatial cognition. (f) shows the subject-wise confusion rate of LMS processing to language processing and spatial cognition, respectively. The error bar denotes the standard error (SE). The asterisks denote the significance of the difference between the confusion rates. *$p$<.05, **$p$<.01, ***$p$<.001.

The observed anatomical and functional similarities between LMS processing and spatial cognition lead to an interesting question: Are the LMS system and the spatial cognition system distinct entities, with the latter supporting the former (Figure 4a)? Alternatively, could it be that the LMS processing is essentially one of the many manifestations of spatial cognition, rendering them indistinguishable in terms of neural correlates (Figure 4b)? To test these two hypotheses, we calculated the intra-LMS similarity by averaging pairwise pattern similarities among LMS tasks (i.e., reasoning, calculation, and mental programming) and the LMS-spatial similarities by averaging the similarities between each LMS task and the representative spatial cognition task (i.e., mental folding) based on their multivariate pattern in the ROI. The underlying rationale is that if the LMS processing is indeed synonymous with spatial cognition, LMS-spatial similarity should fall within the range of intra-LMS similarity. We also calculated LMS-language similarity by averaging the similarities between each LMS task and the representative language task (i.e., sentence processing) in the ROI, and expected it to be lower than both intra-LMS and LMS-spatial similarity.

As expected, the LMS-language similarity is significantly lower than both the intra-LMS similarity ($t(59) = 5.978, p < 0.001$, *Cohen's d* = 0.778) and the LMS-spatial similarity ($t(59) = 7.255, p < 0.001$, *Cohen's d* = 0.945). Critically, the intra-LMS similarity did not significantly exceed the LMS-spatial similarity (one-tail $t(59) = -1.911$, $p = 0.061$, *Cohen's d* = -0.249), challenging the hypothesis that LMS processing and the spatial cognition are distinct entities. This point is further supported by the hierarchical clustering analysis on the multivariate activation patterns of the LMS and spatial cognition tasks. As shown in Figure 4c, the LMS tasks did not form a distinct cluster separate from the spatial cognition task, revealing an intermingled relation among these tasks. This finding supports the idea that LMS processing likely operates via the same underlying mechanisms as spatial cognition.

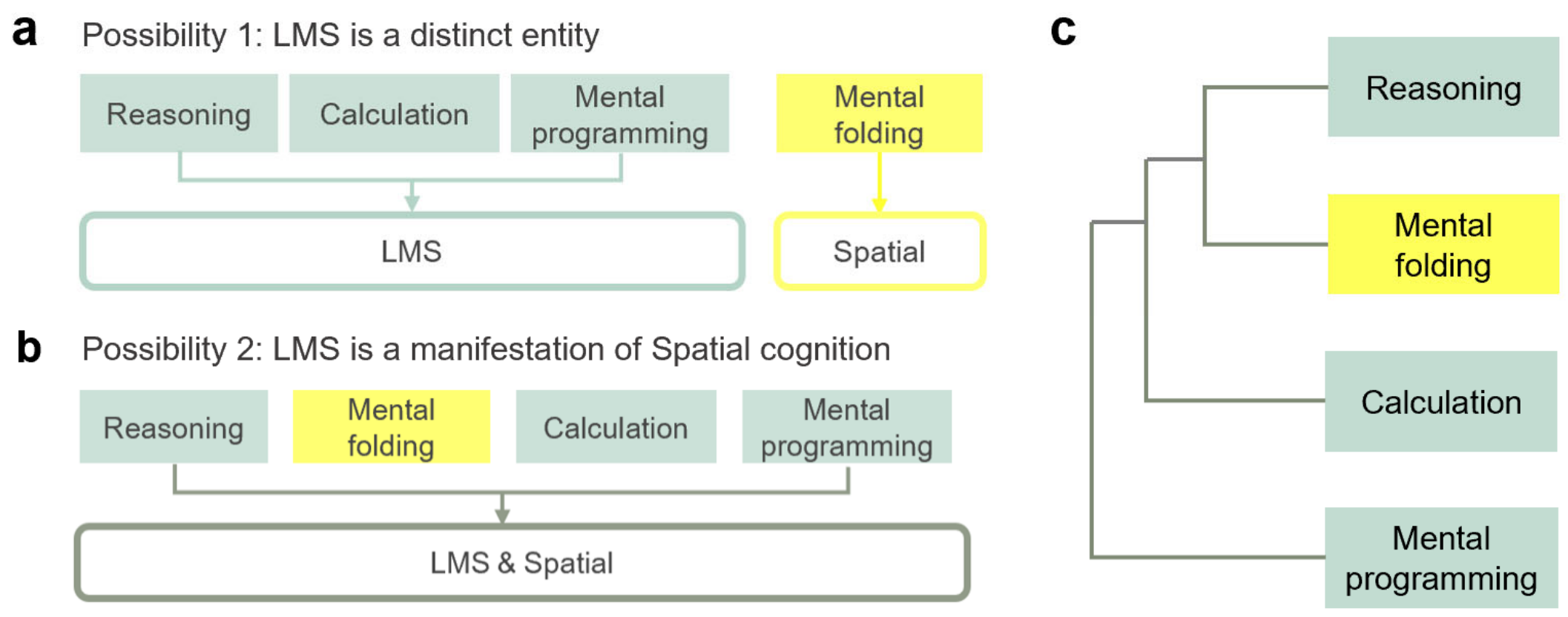

Figure 4. The multivariate pattern similarity between individual LMS processing tasks and spatial cognition. (a) Possibility 1: the LMS processing and the spatial cognition system are distinct entities. (b) Possibility 2: the various LMS tasks and spatial tasks are essentially manifestations of the same system. (c) The hierarchical cluster of the multivariate activation patterns in the ROI. The longer horizontal line illustrates a higher distance and later inclusion in the cluster.

# 3. Discussion

The ability to process logical-mathematical symbols underpins everyday human activities such as calculation, reasoning, and programming. Based on the neuronal recycling hypothesis (Dehaene & Cohen, 2007), we investigated the functional origin of this processing by comparing the neural correlates of LMS processing with those of language processing and spatial cognition, respectively. The comparison of spatial extents in meta-analysis maps revealed a more substantial overlap between LMS processing and spatial cognition than that between LMS and language processing. Furthermore, a pattern similarity analysis revealed that the multivariate activation patterns of LMS processing were more similar to that of spatial cognition than language processing in regions activated by both. Taken together, our findings support the hypothesis that the ability of LMS processing is likely built upon spatial cognition rather than language processing.

The present study presented a novel and direct comparison of the cross-task neural correlates of LMS processing with its two hypothetical origins. The present study differed from previous studies on neural commonality between LMS and spatial processing tasks (Endres, Karas, et al., 2021; Hawes & Ansari, 2020; Huang et al., 2019; Xu et al., 2021; Zago et al., 2008) in overcoming task idiosyncrasy by meta-analysis and directly comparing competing hypotheses to address the theoretical question. First, by utilizing meta-analysis we overcome the impact of task idiosyncrasy which limited domain-level conclusions via task-wise comparison to mitigate the confounding factors. LMS tasks are integrative and idiosyncratic, given the large range of cognitive challenges they target. For instance, our previous study (Xu et al., 2021) observed dissociable components within the cortical regions activated during mental programming, a representative LMS task, each of which showed a varied combination of neural resemblance with another three representative LMS tasks. Also, their resemblance with that of mental folding varied. This highlighted the substantial task idiosyncrasy in LMS processing, which leads to difficulty in evaluating LMS's neural resemblances with other domains, let alone comparing its relative resemblance with multiple other domains. Out of this consideration, the present study compared the cross-task neural correlates of the domains in question via either meta-analysis or synthesized activation map to mitigate the confounding factor of the task-specific properties in identifying the neural correlates.

Specifically, the cross-task map of LMS processing was generated by synthesizing in-house fMRI data of representative LMS tasks with image-based meta-analysis (Salimi-Khorshidi et al., 2009; Soares et al., 2016). The resulting map revealed bilateral activation, albeit slightly right-lateralized, in the frontal lobe (the middle frontal gyrus, the inferior frontal gyrus), the supplementary motion area spanning to the cingulum, the insula, the posterior parietal cortex, the precuneus, and the inferior temporal gyrus. This map covers most major regions previously reported in LMS tasks, including findings from meta-analyses in mathematics (Hawes et al., 2019; Hawes & Ansari, 2020; Tablante et al., 2023), reasoning (Hobeika et al., 2016; Wang et al., 2020), problem-solving (Bartley et al., 2018; Feng et al., 2021), and programming (Castelhano et al., 2021; Ikutani, Itoh, et al., 2021), among other representative LMS tasks (Floyd et al., 2017; Houdé & Tzourio-Mazoyer, 2003; Huang et al., 2019; Ikutani, Kubo, et al., 2021; Y.-F. Liu et al., 2020; Maruyama et al., 2012). By comparing this map with the meta-analysis map of language and spatial processing, we observed that the overlap between LMS processing and spatial cognition mainly resided in a bilaterally distributed set of frontoparietal regions, which are implicated in tasks such as arithmetic calculation, formal logical inference, mental programming, mental rotation, mental folding, and spatial working memory (Hawes et al., 2019; Ivanova et al., 2020; Y.-F. Liu et al., 2020; Xu et al., 2021; Zago et al., 2008). It appears that these regions might be the sites of the reuse of mechanisms for processing objects and their spatial relations. For instance, the frontoparietal dorsal attention network (Corbetta & Shulman, 2002; Kincade et al., 2005) may contribute to maintaining (Ravizza et al., 2004; Sack, 2009) and representing (Hanakawa et al., 2002; Pinel et al., 2001) both spatial codes and LMS, the right insula may be involved in decision-making and error monitoring in both domains (Castelhano et al., 2021; Uddin, 2015), and the inferior parietal lobule may be involved in retaining (Marshuetz et al., 2000) and reactivating (Corbetta et al., 2002) ordered information in both spatial and non-spatial domains, especially in processing programming algorithms or abstract rule sets (Crittenden et al., 2016; Ikutani, Kubo, et al., 2021; Pischedda et al., 2017; Srikant et al., 2022; Woolgar et al., 2011). In contrast, the overlap in neural activation between LMS and language processing was significantly smaller, mainly localized in the left inferior frontal gyrus, including Broca's area (BA45 and BA44). The involvement of these regions in LMS processing may result from task characteristics, such as the text-based visual representation of stimuli presented in LMS tasks (Castelhano et al., 2021). In addition, while traditionally regarded as language-specific regions, recent studies have shown that the BA44 also engages in non-verbal working memory (Gould et al., 2003; Honey et al., 2000; Kong et al., 2005; Maguire et al., 1997), and the BA45 in retrieval and selection of non-linguistic information (Liakakis et al., 2011; Thompson-Schill et al., 1999). The finding of limited LMS-language overlap is consistent with observations that LMS systems emphasize repetition and symmetry, contrasting natural languages that tend to avoid repetition and rely on anti-symmetry. That is, despite both LMS and language utilizing discrete symbol systems, they differ fundamentally in their structural principles (Dehaene et al., 2022). Taken together, the observed neural similarities in this study are in line with the idea that LMS processing likely originates from spatial cognition, possibly through a process of neuronal recycling.

A second distinct feature of the present study is that we achieved a direct comparison and placed the two hypothetical origins in a fair playground by focusing on the regions where both spatial and language processing activate in analyzing multivariate neural resemblance. The results further supported the spatial-origin hypothesis by showing that all tested LMS tasks were carried out in a way more like spatial than language processing when both are plausible. Note that for inferential statistics we leveraged the fMRI data from the same sample of participants in a representative language (sentence processing) task and a spatial (mental folding) task, and calculated the subject-level pairwise similarity among tasks to statistically compare intra-LMS similarity with LMS-spatial and LMS-language similarity. This reanalysis of previously reported data (Xu et al., 2021) was however distinct from previous analyses in the research question, the methodology, the anatomical scope, and the range of tasks included. The previous analysis was conducted within the cortical regions activated by mental programming and broke up the regions in question via clustering analysis based on neural resemblance to a set of non-verbal tasks, while the present study conducted ROI-wise multivariate pattern similarity analysis to compare the relative resemblance of LMS tasks with language and spatial tasks, which could not be predicted by the findings of the previous study.

Though our analysis centered on LMS processing, we indeed speculate that spatial cognition might be a key cognitive cornerstone extending far beyond LMS processing, potentially influencing a wide array of cognitive abilities. Theories like the mental model (Johnson-Laird, 2001) and mental simulation (Barsalou, 2008; Hawes & Ansari, 2020) suggest that relation processing typically utilizes a visuospatial workspace or simulations to navigate representational spaces, and map-like representations are proposed to be a universal coding mechanism that organizes spatial, non-spatial, and even abstract information, using shared neural substrates (Behrens et al., 2018; Igarashi et al., 2022; Theves et al., 2019). For example, the inferior parietal cortex, also identified in this study, is known to encode non-spatial abstract information (Crittenden et al., 2016; Gottlieb & Snyder, 2010; Ikutani, Kubo, et al., 2021; Pischedda et al., 2017;

Srikant et al., 2022; Viswanathan & Nieder, 2020; Woolgar et al., 2011) in addition to spatial information. This suggests a shared neural basis for spatial and non-spatial information mapping. Indeed, the spatial navigation system has been proposed to provide critical computational components for the neural implementation of the "language of thought" (LoT) (Frankland & Greene, 2020; Kazanina & Poeppel, 2023). LoT, advanced by Jerry Fodor (1975), describes not an actual language but a formal symbol system that stores and manipulates concepts symbolically, adhering to compositional rules. This system enables the construction of complex thoughts from primitive symbolic concepts and thus gives rise to the productivity and creativity of the human mind, advocating LoT as a candidate for human singularity in cognition (Dehaene et al., 2022; Sablé-Meyer et al., 2021).

However, the proposal that spatial cognition serves as a cross-domain computational foundation in the human mind leads to a puzzle: why does spatial cognition, an ability shared with many species, uniquely generate sophisticated non-spatial abilities in humans, such as formal operations of LMS systems? One possible answer is that spatial cognition may not be the exclusive origin of LMS processing. Language processing, although not the primary source, may also contribute to LMS processing (Dowker et al., 2008; Ikutani, Itoh, et al., 2021; Wang et al., 2020). Our study relies predominantly on anatomical overlaps in activation maps. This can result in an oversight of functional similarities between different anatomical regions. According to the multiple parallel circuit hypothesis (Dehaene et al., 2022), anatomically separate regions may instead undertake similar functionalities in different symbol systems, which were all based on the recursive composition of corresponding primitives to encode nested repetitions with variations. Thus, future studies should explore the potential computational commonalities between language and LMS processing to further understand these intricate cognitive interactions.

Another limitation of our study is its reliance on correlational evidence, instead of establishing a causal link between LMS processing and spatial cognition. Recent advancements in large language models (LLM) may offer a novel avenue for experimentation on this topic. By “re-running” evolution, we can observe how spatial cognition scaffolds the emergence of LMS processing during the evolution. Supporting this intuition, state-of-the-art LLMs, proficient in natural language tasks, show brittle performance in LMS tasks (Frieder et al., 2023; H. Liu et al., 2023). Wong et al. (2023) suggested that this might be because LLMs work upon the context-aware mapping between language and meanings, lacking a world model based on probabilistic LoT. Based on the link between LMS processing and spatial cognition observed in this study, it could be worth investigating the enhancement of LMS processing in LLMs through training on spatial cognition tasks, such as path integrating, orienting, and map reading. Indeed, recent studies have demonstrated training navigation in abstract concept space using cognitive maps could improve the reasoning and problem-solving ability of LLMs (Stöckl et al., 2024; Sun et al., 2023). Extending such investigation to training spatial cognition and endowing LLMs with spatial abilities might bridge the gap between disembodied information in LLM training and embodied reality that artificial general intelligence (AGI) ultimately aims to interact with, which could significantly contribute to the development of next-generation AGI.

# 4. Methods

## 4.1. Meta-analyses of language processing and spatial cognition

We generated meta-analysis activation maps based on large-scale automated meta-analyses of functional magnetic resonance imaging (fMRI) for language processing and spatial cognition.

### 4.1.1. Meta-analysis of language processing

To generate a map of the neural correlates of language processing, we queried the Neurosynth (Yarkoni et al., 2011) database (version 0.5 released on February 23, 2015) using the term "language" (accessed on 2023.8.19). This yielded a collection of 823 fMRI studies that exhibited significant relevance to the term with the default frequency threshold of 0.001. Based on the coordinate-based meta-analysis technique of Neurosynth, we generated the map of brain regions that were preferentially engaged across the selected studies. Each voxel within the resulting map was labeled by the z-score resulting from a two-way ANOVA assessing the concomitant presence of term loading and voxel activation. The map was further corrected for multiple comparisons employing the false discovery rate (FDR) approach with an anticipated FDR threshold of 0.01.

### 4.1.2. Meta-analysis of spatial cognition

As for spatial cognition, there are no uniform terms to search in Neurosynth. Therefore, we used the BrainMap database to generate the meta-analysis map of spatial cognition. Following the taxonomy outlined by Fox et al. (2005), we curated the experiments categorized under the Behavioral Domain "Cognition.Spatial" in the BrainMap functional database (accessed by Sleuth 3.0.4 in 2023.8.19), under the constraint of "activations only". This curation yielded 251 experimental contrasts from 63 relevant publications. The meta-analysis based on the experimental contrasts was implemented by the activation likelihood estimation (ALE), with BrainMap's GingerALE tool. Same as the language processing analysis, we applied FDR correction (at $p = 0.01$) to the resulting z-score map, adhering to the default GingerALE settings.

## 4.2. fMRI studies

The present study used fMRI data from a separate sample of participants to estimate the neural correlates of a set of representative LMS processing tasks, in addition to a representative language processing task and a representative spatial cognition task. Analyses based on these data for a different purpose have been reported in a previous study (Xu et al., 2021).

### 4.2.1. Participants

Twenty participants (right-handed neurologically normal volunteers with normal or corrected-to-normal vision) completed the experiment. The participants were screened before recruitment to ensure they either had taken programming-related courses or had experience in programming. The targeted sample size (20 valid participants) was decided a priori, to be similar to previous studies using similar tasks (e.g., Fedorenko et al.(2012), n = 13-16; Zhou et al.(2018), n = 24; Amalric & Dehaene(2016), n = 15). The study was approved by the Institutional Review Board of BNU. Written informed consent was obtained from all participants before they took part in the experiment, and the participants received money for their time.

For a full description of the fMRI data acquisition, see Xu et al. (2021). For each participant, the fMRI data of three LMS processing tasks, one language processing task, and one spatial cognition task were recorded and analyzed in the present study.

### 4.2.2. Experimental paradigms

**LMS processing**. BOLD signals during three representative LMS processing tasks, namely mental programming, reasoning, and calculation, were collected.

The mental programming task engaged participants in mentally solving simple programming problems, with reporting familiarity with knowledge entries as the control condition. The programming problems were presented in Chinese as a one-to-three-line introductory statement, a programming probe ("Please design a program with:"), a one-line statement specifying the potential input, and a one-line statement describing the desired output. The knowledge entries in the control condition were matched with the materials of the programming condition in length, format, and the use of symbols. Each entry started with a one-to-three-line introductory statement, followed by a brief conjunctive clause (e.g., "in other words"), and two one-to-two-line elaborations of the concept introduced in the introductory statement. The mental programming trials started with a programming task probe presenting “report progress”, and the participants were to read and consider the programming problem, and wait for the response prompt to choose from a four-point scale to report their progress in programming. The participants were instructed before the scanning that for programming they should consider the algorithm instead of specific code, and the solution of the problem means finding the algorithm instead of finishing the coding. The control trials started with the control task probe which read “report familiarity”, and participants were to read and consider whether they were familiar with the content of the knowledge entry and choose from a four-point scale to report their familiarity with the content. The mental programming task was divided into six runs, each consisting of four trials in the mental programming condition and four in the control condition.

The reasoning task comprised a reasoning condition adopted from Raven’s APM and a control condition. In both conditions, the participants first saw a problem with a complex main figure, with eight choices below. The problems of the reasoning trials were chosen from Raven’s APM, and those of the control condition were composed by rearranging the choices of other Raven’s APM items in the main figure and the choices in each problem. Above each problem, a task probe would indicate whether it is a reasoning trial (“Look for the rule”) or a control trial (“Look for the choices”). The participants were to consider which choice complies with the rule of figure arrangement and completes the main pattern (the reasoning condition), or to consider which choice was presented in the main pattern (the control condition). Each participant completed five functional runs consisting of eight trials each.

The calculation task was adopted from Fedorenko et al. (2010) with an additional control condition. In the calculation trials, the participants would first see a task probe ("Do the addition") and then see one number (11–30) followed by three sequentially presented addends. Then the participants had to choose the correct sum in a two-choice, forced-choice question. In the control trials, after the task probe ("Memorize the numbers") and the same sequential presentation of four numbers, the participants were to select, from two alternatives, the one among the four preceding numbers. Each participant completed four runs which each consisted of six calculation blocks and six control blocks.

**Language processing and spatial cognition.**

The sentence processing task was used as the representative language processing task. It was adopted from Fedorenko et al. (2010) but used sentences and pseudo-characters in Mandarin Chinese. Each experimental trial consisted of a sequential presentation of a string of seven Chinese real words which formed one sentence followed by a target word. The task was to decide whether the target word appeared in the preceding string. In control trials, the Chinese characters were replaced by pseudo-characters. The trials were grouped into six-trial blocks. Each participant underwent four block-design runs, each consisting of three experimental blocks and three control blocks.

Spatial cognition was measured in a mental folding task adopted from Milivojevic et al. (2003). In each trial, the participants were shown a black outline of six squares joined together, representing the unfolded surface of a cube. Two sides belonging to different squares (mental folding condition) or the same square (control condition) were pointed by two small arrows. In the "match" trials, the pointed sides would meet if the squares were folded up into a cube. Trials in the mental folding and the control conditions differed in the total number of squares carried along for each fold for the match–mismatch decision to be made. The control trials presented 1-square-carried stimuli since the judgment of such stimuli can be made without mental folding. Every four trials of the same condition were grouped into a block. Each participant completed five or six block-design functional runs. Note that the first three participants completed six runs, and the rest participants underwent only five runs due to the limitation of the scanning time.

For a complete description of the experimental paradigms, see Xu et al. (2021).

### 4.2.3. fMRI data acquisition and analysis

**Image acquisition.** The fMRI data were acquired using a Siemens 3T scanner with a 12-channel phased-array head coil. Task-state fMRI (ts-fMRI) was acquired using a T2 *-weighted echo-planar-imaging (EPI) sequence with a whole-brain protocol (TR = 2000 ms, TE = 30 ms, flip angle = 90◦, and in-plane resolution = 3.1 × 3.1 × 3.5 mm, 33 contiguous interleaved slices). For a complete description of the imaging parameters, see Xu et al. (2021).

**Preprocessing.** The functional data were preprocessed with DPABI (Yan et al., 2016). The main preprocessing procedure included slice timing, head motion correction, co-registration, segmentation, spatial normalization to the standard MNI space, and resampling to 3 × 3 × 3 mm isotropic voxels. The data were smoothed with a 4 mm FWHM kernel for univariate activation analysis. Smoothing was omitted for multivariate activation analysis. Runs characterized by excessive head movement were excluded from further analysis.

**Univariate analyses.** First-level analyses were performed for functional images of each participant in each task using the general linear model (GLM). The GLM incorporated regressors corresponding to the following events: the task condition, the control condition, and the presence of the response prompt and task probe. The regressors were convolved with the canonical hemodynamic response function (HRF). A 1/128 Hz high-pass filter was applied to remove low-frequency noise, with the AR (1) model used to account for serial correlations. For each experiment, subject-level and group-level analyses were conducted within a gray matter mask by SPM8 (Wellcome Department of Imaging Neuroscience, London; www.fil.ion.ucl.ac.uk/spm). The subject-level map corresponded to the contrasts of parameter estimates (COPE) between the respective task versus the control condition of each subject. The group-level map was estimated in the second-level analysis of the corresponding task-control contrasts.

#### 4.2.4. Conjunctive analysis of LMS processing

To estimate the neural correlates across diverse LMS processing tasks, we generated a synthesized activation map of LMS processing based on the data of the three representative LMS tasks, i.e., reasoning, calculation, and mental programming tasks described above. We utilized Stouffer's method in the image-based meta-analysis. The unthresholded group-level maps of each LMS processing task were subjected to Stouffer's estimator using the NiMARE Python package version 0.0.13 (Salo et al., 2023). The estimator combined the group-level maps via Stouffer's method (Stouffer et al., 1949) and produced a fixed-effect estimate of the combined effect, to obtain the most sensitive pooling of the studies. Each voxel within the resultant synthesized map conveyed a third-level COPE, indicating the pooled activation across LMS processing tasks. The synthesized map was corrected with an expected FDR of 0.01.

### 4.3. Dice coefficients of meta-analysis maps and subject-level maps

To quantify the extent of overlap between LMS processing and language processing and that between LMS processing and spatial cognition, we computed dice coefficients between the activation maps for each pair, which was an index of the convergence between activation maps. For a given pair of maps, the dice coefficient was determined as the ratio between twice the count of shared voxels and the sum of voxels in both maps. The dice coefficients were calculated between the thresholded meta-analysis map of language processing and the thresholded synthesized map of LMS processing, and between spatial cognition and LMS processing. In addition, for statistical comparison, corresponding dice coefficients were calculated for each subject based on the corresponding subject-level maps (thresholded by a voxel-level threshold at $p = 0.001$).

### 4.4. Multivariate pattern analysis

To compare the multivariate pattern similarity between LMS processing and those associated with language processing and spatial cognition, we conducted multivariate pattern analysis in an ROI-based manner. The region of interest (ROI) was defined as the common activation region between the meta-analysis maps of language processing and spatial cognition.

The comparison of multivariate activation pattern similarity was quantified by the confusion rate in multivariate activation pattern classification. Specifically, we trained a classifier for each subject to classify the multivariate activation pattern in the run-level beta map of the sentence processing task from that of the mental folding task. The run-level beta map (task versus fixation) was based on the GLM using non-smoothed data. The classifier was based on the support vector machine (SVM) with a linear kernel implemented by The Decoding Toolbox (TDT) (Hebart et al., 2015), and trained with the leave-one-out cross-validation (LOOCV) to prevent overfitting. In each cross-validation step, one SVM was trained by three runs of sentence processing task and three runs of mental rotation task separated from one run of each task kept unknown. Then in the test session, the SVM would classify the left-out run from each training task for performance validation (indicated as accuracy). For quantitative balance, all tasks included the first four runs in a set. Note that the outcome appeared consistent regardless of which specific four runs were chosen in the examination. The LOOCV traversed all combinations of left-out runs and resulted in an averaged classifier of each participant. The classifier achieved an accuracy of nearly 100% in distinguishing the two tasks across 20 participants. In the test session, each run of LMS processing tasks was classified by the SVM into one of the training tasks. The confusion rate at which LMS processing was (mis)classified as a given training task quantified the pattern resemblance between the LMS processing task and the corresponding training task.

To test the relationship between LMS processing and spatial cognition, we calculated the intra-LMS similarity, LMS-spatial similarities, and LMS-language similarity based on the same run-level beta maps in the ROI. For each participant, the multivariate pattern for a representative task was represented as a one-dimensional vector concatenated sequentially from the first four runs' vectors. The pairwise multivariate pattern similarities were calculated using Spearman's r for all corresponding combinations of two tasks for each participant and statistically tested at the group level.

To further investigate the hierarchical multivariate pattern similarity between LMS processing and spatial cognition, agglomerative hierarchical clustering analysis was conducted on the multivariate patterns of the tasks, based on the same run-level beta maps in the ROI. For each representative task, the multivariate pattern was a one-dimensional vector formed by concatenating the multivariate pattern from each participant. The clustering analysis was conducted with the nearest point algorithm implemented by the SciPy Python package version 1.5.4 (Virtanen et al., 2020).

## 5. Acknowledgments

This study was funded by Natural Science Foundation of China (2371099), Beijing Municipal Science & Technology Commission, Administrative Commission of Zhongguancun Science Park (Z221100002722012), and Double First-Class Initiative Funds for Discipline Construction.

# Supplementary materials

## 7. Supplementary figures

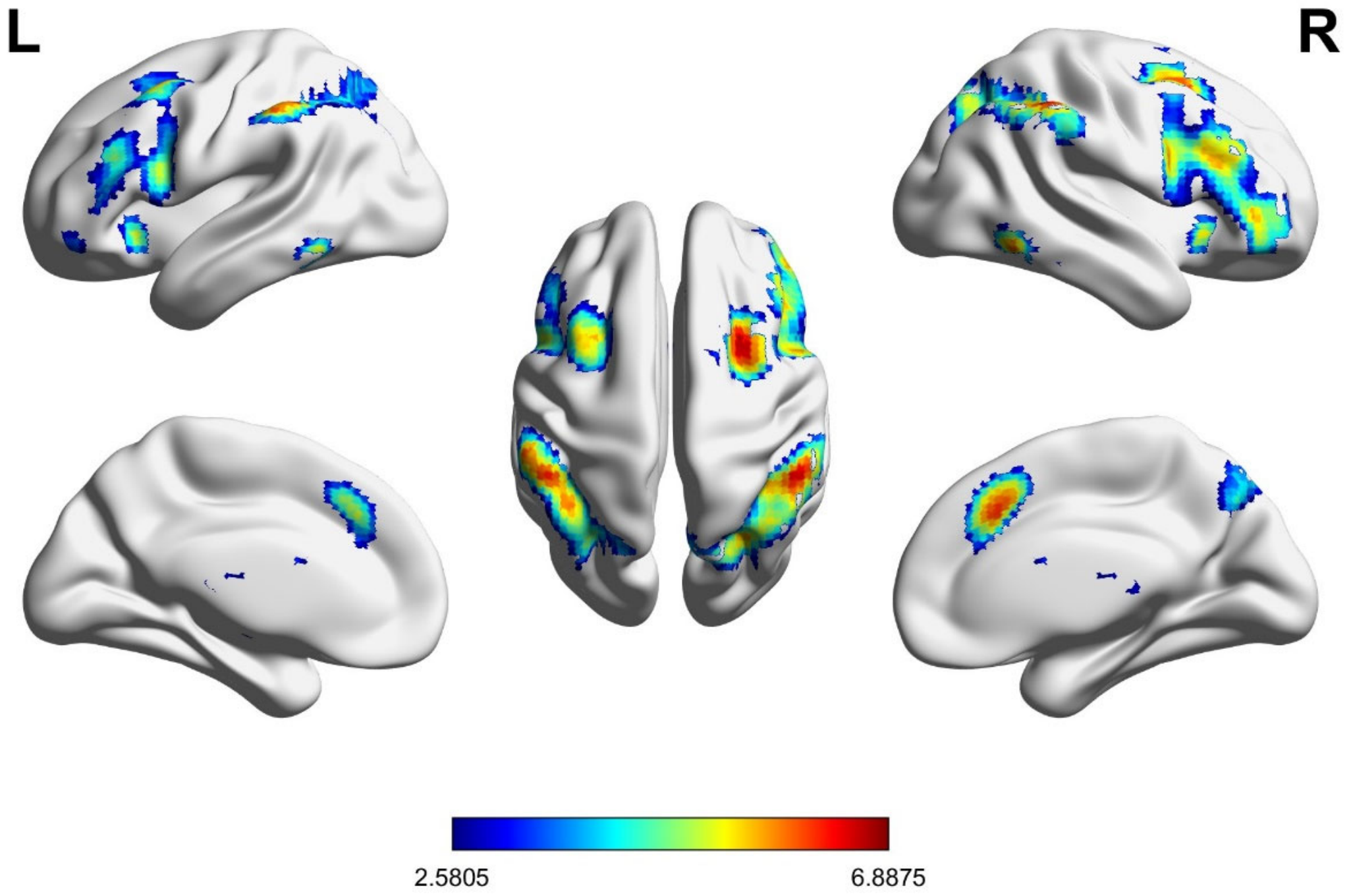

Supplementary Figure 1. Synthesized activation map of LMS processing from conjunctive analysis of three representative tasks. The activation maps were FDR corrected at $p$ = 0.01. The warmer color denotes a higher z-score.

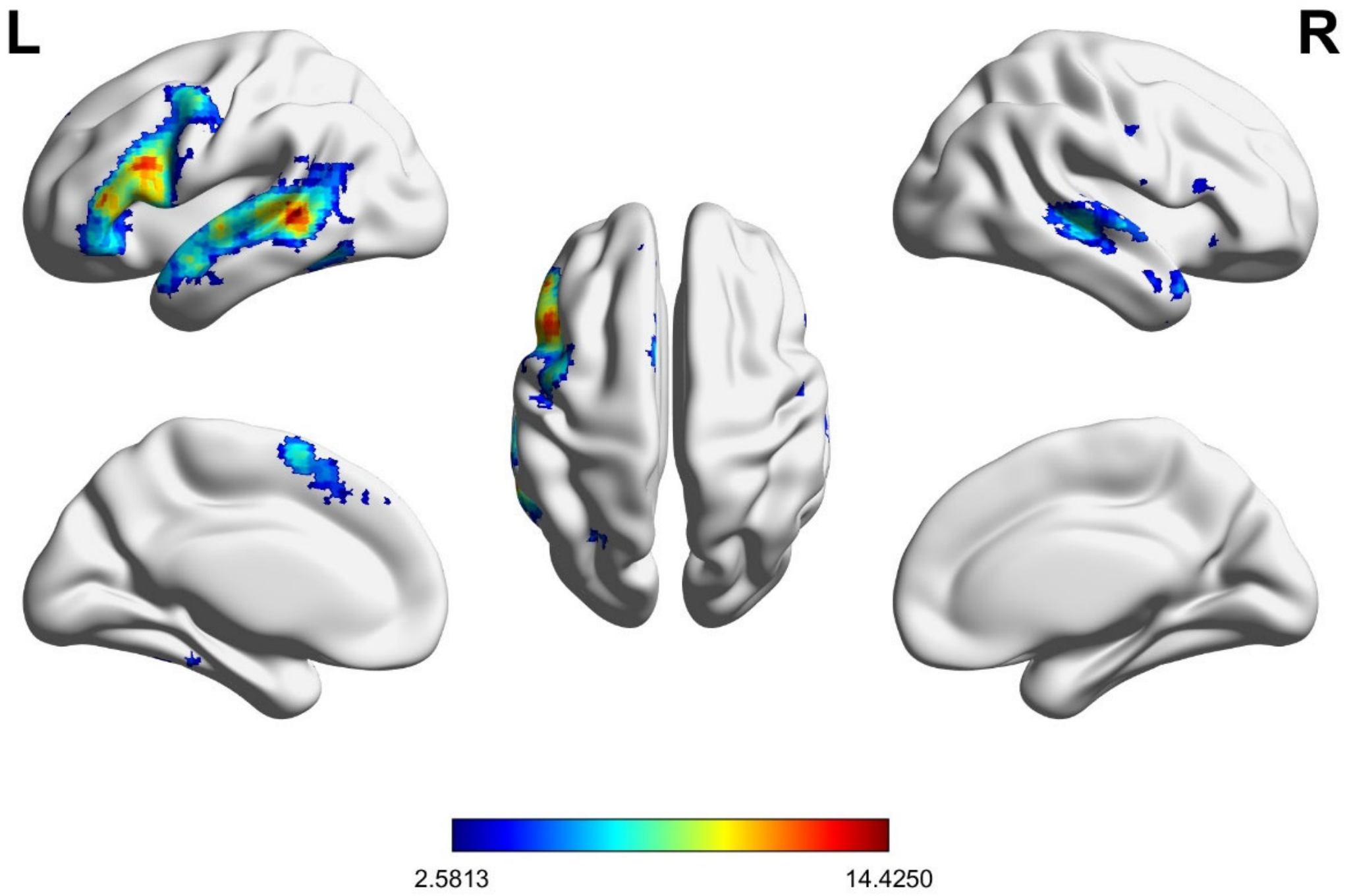

Supplementary Figure 2. Meta-analysis activation map of language processing from large-scale automated meta-analysis. The activation maps were FDR corrected at $p$ = 0.01. The warmer color denotes a higher z-score.

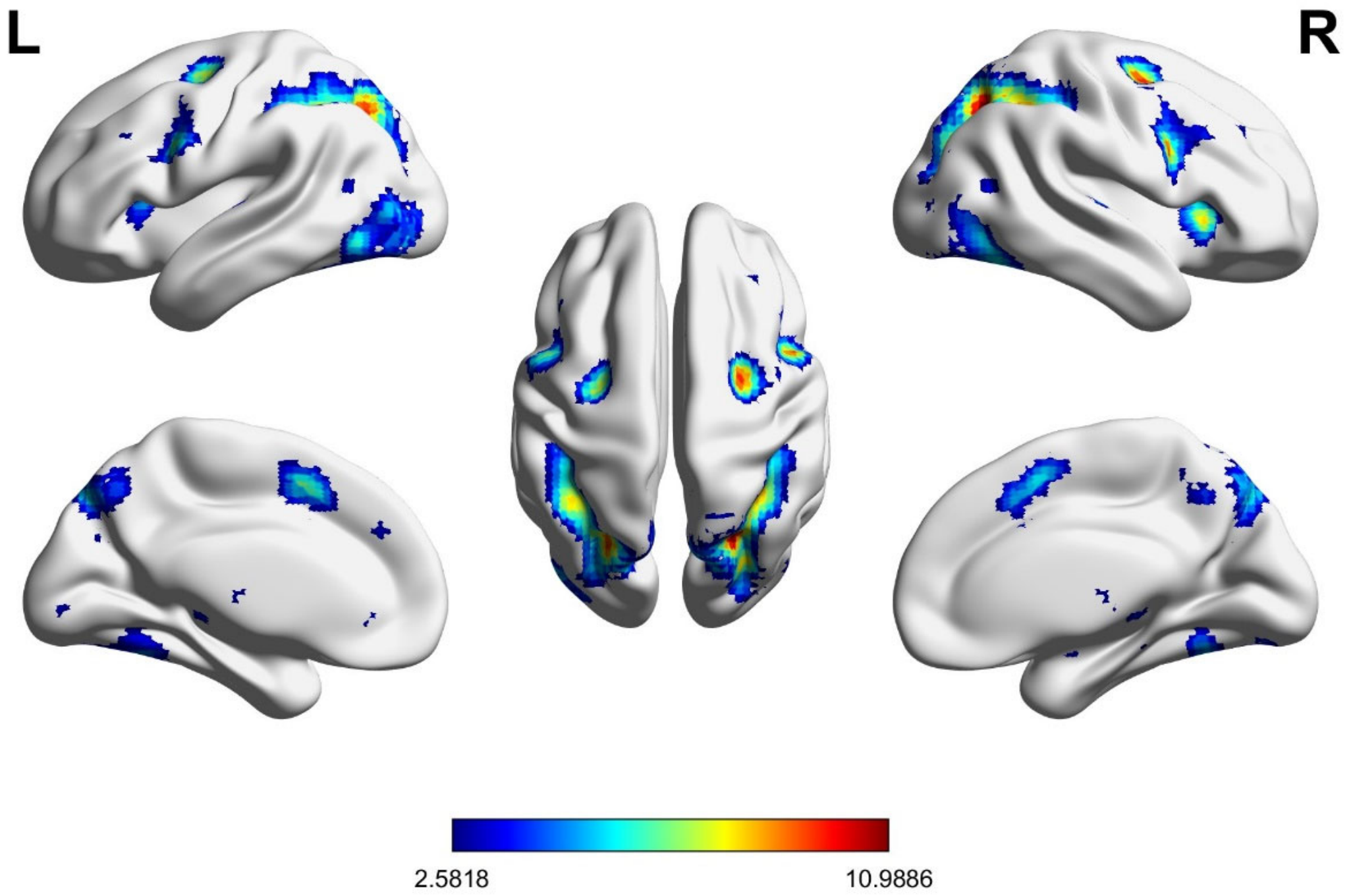

Supplementary Figure 3. Meta-analysis activation map of spatial cognition from large-scale automated meta-analysis. The activation maps were FDR corrected at $p$ = 0.01. The warmer color denotes a higher z-score.

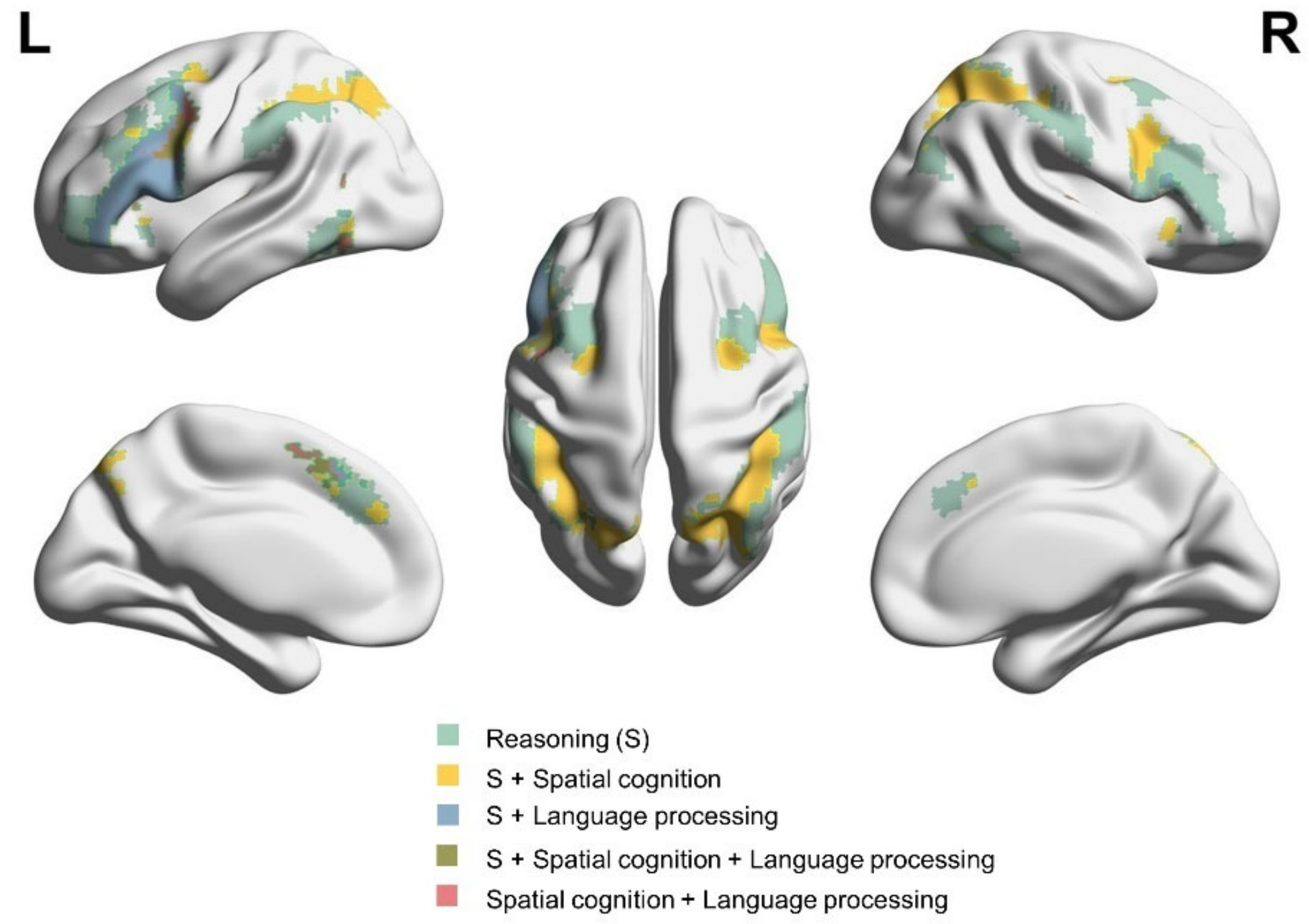

Supplemental Figure 4. Overlaps between the activation map of reasoning and the meta-analysis activation maps of language processing and spatial cognition. All the activation maps were FDR corrected at $p = 0.01$.

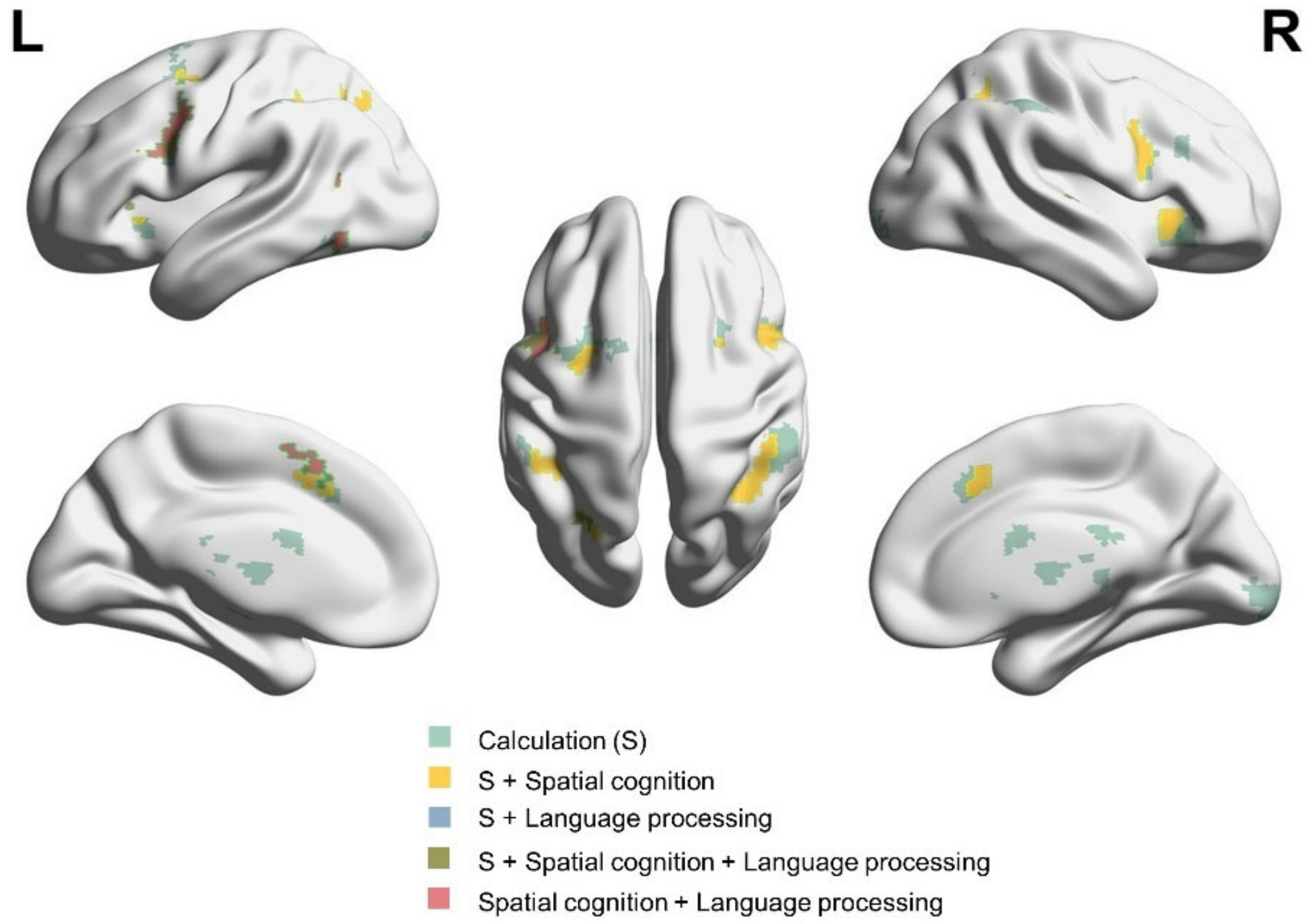

Supplemental Figure 5. Overlaps between the activation map of calculation and the meta-analysis activation maps of language processing and spatial cognition. All the activation maps were FDR corrected at $p$ = 0.01.

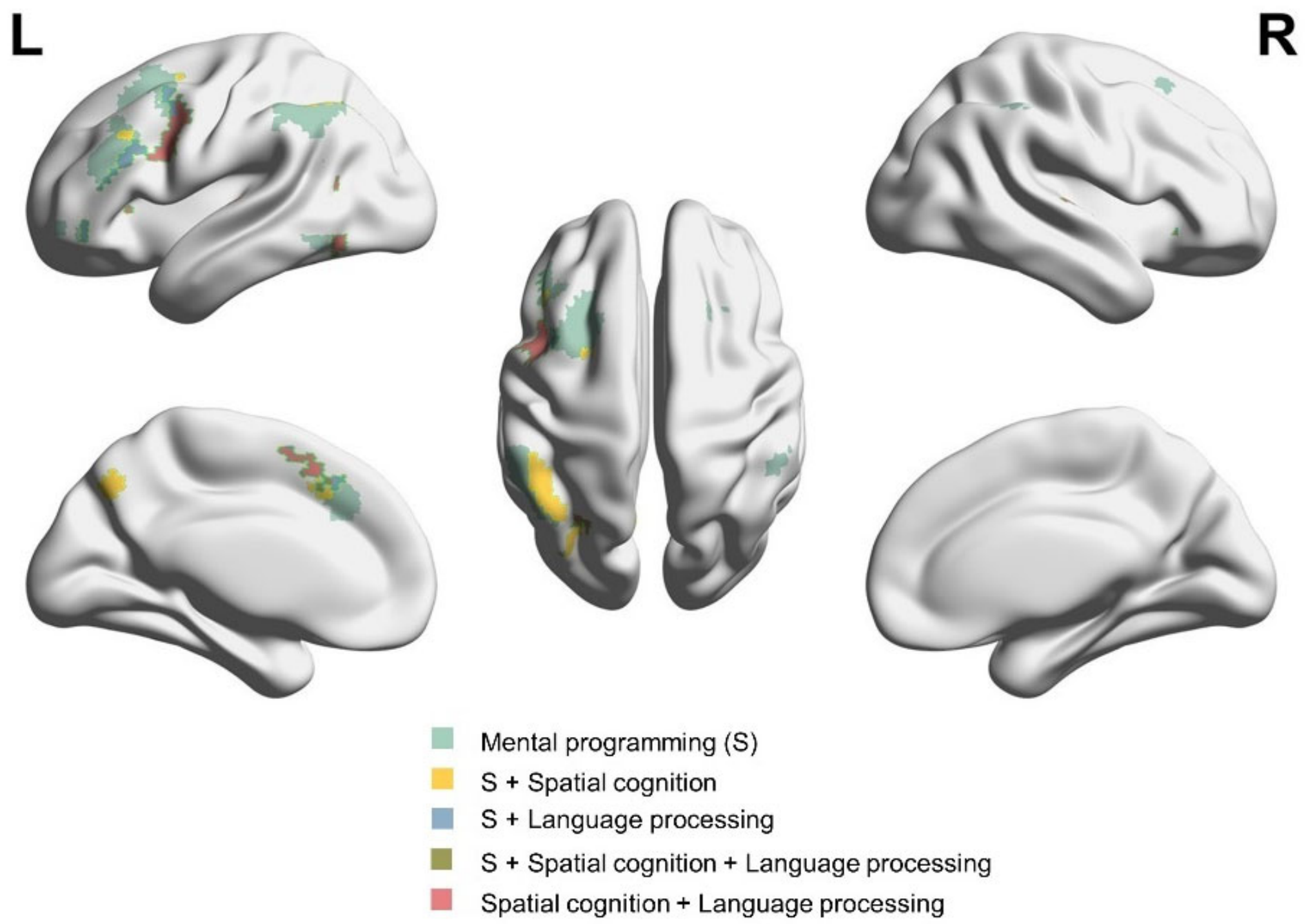

Supplemental Figure 6. Overlaps between the activation map of mental programming and the meta-analysis activation maps of language processing and spatial cognition. All the activation maps were FDR corrected at $p = 0.01$.

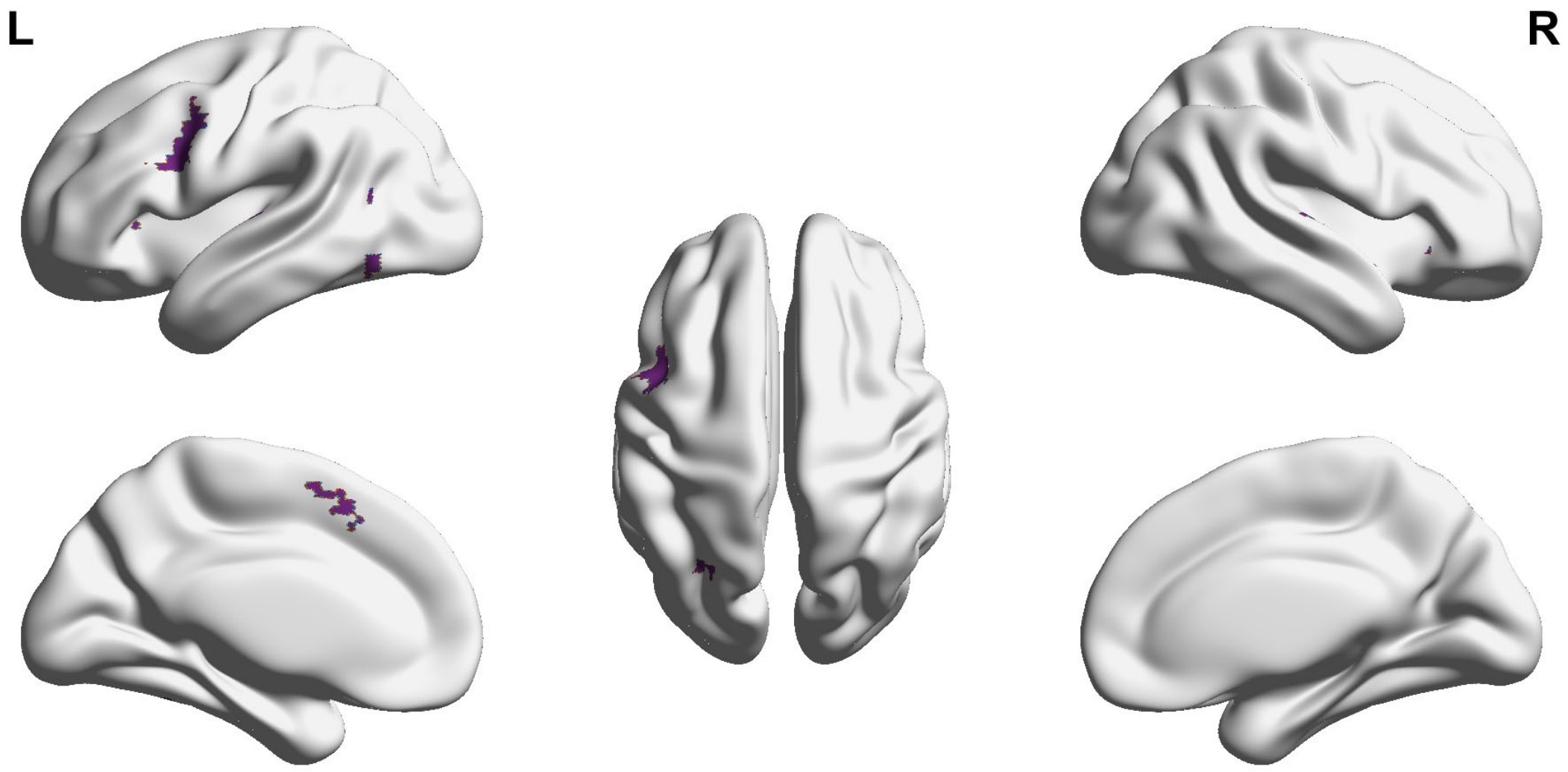

Supplementary Figure 7. The ROI of the multivariate similarity analysis. The ROI is the overlap between the meta-analysis activation maps of language processing and spatial cognition.

## 8. Supplementary tables

Supplementary Table 1 The activation regions of LMS processing.

| Hemisphere | Region | MNI coordinates | | | Volume ($mm^3$) |
|---|---|---|---|---|---|
| | | x | y | z | |
| Left | Inferior Frontal Gyrus, Middle Temporal Gyrus, Superior Temporal Gyrus, Inferior Temporal Gyrus, Superior Temporal Pole, Postcentral Gyrus | -54 | -42 | 3 | 74250 |
| | Inferior Parietal Lobule, Superior Parietal Lobule | -30 | -66 | 39 | 567 |
| | Supplementary Motor Area | -3 | 6 | 60 | 3618 |
| Right | Middle Temporal Pole, Superior Temporal Pole | 48 | 12 | -27 | 1593 |
| | Superior Temporal Gyrus, Middle Temporal Gyrus | 54 | -24 | 0 | 10017 |
| | Inferior Frontal Gyrus | 57 | 21 | 12 | 864 |
| | Postcentral Gyrus | 48 | -6 | 36 | 432 |
| | Precentral Gyrus | 60 | -3 | 42 | 405 |

Supplementary Table 2 The activation regions of language processing.

| Hemisphere | Region | MNI coordinates | | | Volume ($mm^3$) |
|---|---|---|---|---|---|
| | | x | y | z | |
| Left | Cerebellum Posterior Lobe, Cerebellum Anterior Lobe | -30 | -66 | -33 | 40554 |
| | Caudate, Thalamus, Pallidum, Putamen, Globus Pallidus | -15 | 6 | 9 | 9774 |
| | Inferior Temporal Gyrus, Middle Temporal Gyrus | -57 | -51 | -12 | 2295 |
| | Inferior Frontal Gyrus, Middle Frontal Gyrus | -45 | 48 | -9 | 1350 |
| | Orbital part of Inferior Frontal Gyrus, Insula | -33 | 21 | -6 | 2268 |
| | Inferior Frontal Gyrus, Middle Frontal Gyrus, Precentral Gyrus | -30 | 12 | 48 | 27567 |
| | Thalamus | -18 | -27 | 12 | 297 |
| | Inferior Parietal Lobule, Superior Parietal Lobule, Angular, Middle Occipital Gyrus, Supramarginal Gyrus, Postcentral Gyrus | -48 | -42 | 45 | 22950 |
| Right | Cerebellum Posterior Lobe, Brainstem | 18 | -36 | -45 | 1296 |
| | Inferior Temporal Gyrus | 57 | -54 | -12 | 3105 |
| | Inferior Frontal Gyrus, Middle Frontal Gyrus, Superior Frontal Gyrus, Cingulate Gyrus, Supplementary Motor Area, Insula | 27 | 9 | 51 | 60075 |
| | Caudate, Thalamus, Pallidum, Putamen | -33 | 21 | -6 | 2268 |
| | Inferior Parietal Lobule, Angular, Supramarginal Gyrus, Precuneus, Superior Parietal Lobule, Superior Occipital Gyrus, Middle Occipital Gyrus, Cuneus | 51 | -42 | 45 | 28755 |

Supplementary Table 3 The activation regions of spatial cognition.

| Hemisphere | Region | MNI coordinates | | | Volume ($mm^3$) |
|---|---|---|---|---|---|
| | | x | y | z | |
| Left | Inferior Parietal Lobule, Middle Occipital Gyrus, Superior Parietal Lobule, Fusiform, Inferior Occipital Gyrus, Precuneus, Superior Occipital Gyrus, Angular, Postcentral Gyrus, Inferior Temporal Gyrus | -21 | -69 | 45 | 42687 |
| | Thalamus | -9 | -21 | 6 | 2862 |
| | Precentral Gyrus, Insula, Superior Frontal Gyrus, Inferior Frontal Gyrus | -24 | -6 | 54 | 16686 |
| | Middle Occipital Gyrus | -24 | -69 | 3 | 486 |
| | Precentral Gyrus, Superior Temporal Gyrus | -48 | -15 | 12 | 783 |
| | Medial Superior Frontal Gyrus | -6 | 39 | 30 | 270 |
| | Supplementary Motor Area, Cingulate | -3 | 6 | 48 | 10179 |
| | Middle Frontal Gyrus | -36 | 24 | 33 | 405 |
| Right | Middle Occipital Gyrus, Superior Occipital Gyrus, Inferior Parietal Lobule, Superior Parietal Lobule, Precuneus, Fusiform, Inferior Occipital Gyrus, Inferior Temporal Gyrus, Angular, Cuneus, Middle Temporal Gyrus, Supramarginal Gyrus, Postcentral Gyrus | 24 | -66 | 45 | 50679 |
| | Insula, Opercular part of Inferior Frontal Gyrus, Putamen | 30 | 21 | 0 | 4968 |
| | Thalamus, Hippocampus | 15 | -30 | -3 | 810 |
| | Calcarine Sulcus | 3 | -72 | 9 | 810 |
| | Precentral Gyrus, Superior Temporal Gyrus | 48 | -15 | 12 | 1458 |
| | Precentral Gyrus, Superior Frontal Gyrus, Opercular part of Inferior Frontal Gyrus, Middle Frontal Gyrus | 27 | -6 | 54 | 13716 |
| | Middle Frontal Gyrus, Triangular part of Inferior Frontal Gyrus | 33 | 33 | 33 | 594 |

Supplementary Table 4 The overlap of activation regions between the LMS processing and the two compared tasks (i.e., language processing and spatial cognition).

| Compared Task | Hemisphere | Region | MNI coordinates | | | Volume (mm$^3$) |
|---|---|---|---|---|---|---|
| | | | x | y | z | |
| Language processing | Left | Inferior Temporal Gyrus | -51 | -46 | -12 | 297 |
| | | Inferior Frontal Gyrus | -48 | 44 | -7 | 378 |
| | | Inferior Frontal Gyrus, Insula | -33 | 27 | -3 | 378 |
| | | Inferior Frontal Gyrus, Middle Frontal Gyrus, Precentral Gyrus | -48 | 17 | 22 | 13743 |
| | Right | Inferior Frontal Gyrus | 54 | 23 | 18 | 459 |
| | | Inferior Parietal Gyrus, Superior Parietal Gyrus | -28 | -66 | 41 | 540 |
| | | Supplementary Motor Area, Medial Superior Frontal Gyrus | -3 | 22 | 47 | 891 |
| Spatial Cognition | Left | Inferior Frontal Gyrus, Precentral Gyrus | -44 | 9 | 28 | 2565 |
| | | Inferior Parietal Gyrus, Superior Parietal Gyrus, Angular | -35 | -55 | 43 | 12177 |
| | | Middle Frontal Gyrus, Precentral Gyrus | -30 | 1 | 49 | 837 |
| | | Insula | -30 | 21 | -2 | 351 |
| | Right | Brainstem | 0 | -29 | -16 | 405 |
| | | Supplementary Motor Area(L&R), Medial Superior Frontal Gyrus, Middle Cingulate | 3 | 19 | 43 | 2943 |
| | | Inferior Parietal Gyrus, Superior Occipital Gyrus, Precuneus, Angular, Superior Parietal Gyrus, Middel Occipital Gyrus | 28 | -61 | 42 | 15768 |
| | | Inferior Frontal Gyrus, Insula | 31 | 21 | -3 | 1458 |
| | | Inferior Frontal Gyrus | 38 | 29 | 25 | 297 |
| | | Inferior Frontal Gyrus, Middle Frontal Gyrus, Precentral Gyrus | 40 | 5 | 37 | 8127 |
| | | Inferior Temporal Gyrus | 48 | -58 | -10 | 270 |

Supplementary Table 5 The ROI of the multivariate similarity analysis.

| Hemisphere | Region | MNI coordinates | | | Volume ($mm^3$) |
|---|---|---|---|---|---|
| | | x | y | z | |
| Left | Inferior Occipital gyrus | -45 | -61 | -12 | 351 |
| | Superior Temporal gyrus | -52 | -19 | 5 | 270 |
| | Precentral Gyrus, Inferior Frontal lobule | -44 | 7 | 30 | 4644 |
| | Inferior Parietal gyrus, Superior Parietal gyrus | -29 | -63 | 43 | 972 |
| | Supplementary Motor Area | -4 | 13 | 52 | 1431 |
| Right | Superior Temporal Gyrus | 52 | -21 | 4 | 405 |